\documentclass[12pt]{article}
\usepackage{amssymb}
\usepackage{cite}
\usepackage{graphicx}
\usepackage{epsfig}

\begin{document}

\title{From Diffusion to Anomalous Diffusion: A Century after Einstein's Brownian Motion}
\date{\today}

\author{I.M. Sokolov\\
Institut f\"{u}r Physik, Humboldt-Universit\"{a}t zu Berlin,\\
Newtonstr. 15, D-12489 Berlin, Germany \\
and\\
J. Klafter\\
School of Chemistry, Sackler Faculty of Exact Sciences,\\
Tel Aviv University, Tel Aviv 69978, Israel}
\maketitle

\begin{abstract}
Einstein's explanation of Brownian motion provided one of the cornerstones which
underlie the modern approaches to stochastic processes. His approach is based on a random walk 
picture and is valid for Markovian processes lacking long-term memory. 
The coarse-grained behavior of such processes is described by the diffusion equation. 
However, many natural processes do not possess the Markovian property and exhibit 
to anomalous diffusion. We consider here the case of subdiffusive processes, which are
semi-Markovian and correspond to continuous-time random walks in which the waiting 
time for a step is given by a probability distribution with a diverging mean value. 
Such a process can be considered as a process subordinated to normal diffusion under 
operational time which depends on this pathological waiting-time distribution. 
We derive two different but equivalent forms of kinetic equations, which reduce 
to know fractional diffusion or Fokker-Planck equations for waiting-time distributions 
following a power-law. For waiting time distributions which are not pure power laws one 
or the other form of the kinetic equation is advantageous, depending on whether the 
process slows down or accelerates in the course of time. 
\end{abstract}

\begin{bf}
The description of Brownian motion in Einstein's 1905 work relies
on the assumption of the existence of a time-interval $\tau$, such 
that the particle's motion during different  $\tau$-intervals is independent.
The coarse-grained version of this motion leads then to the known diffusion equation. 
However, in many cases this assumption is violated. An example is dispersive transport 
in disordered systems which stems from a broad distribution of waiting times which 
may have a diverging mean. This ill-defined mean waiting time results in subdiffusion. 
In this contribution we derive, within a unified scheme, two equivalent forms of 
kinetic equations for subdiffusive behavior. For power-law waiting-time 
distributions, the equations reduce to the "normal" form of a fractional Fokker-Planck 
equation with a fractional derivative replacing the first-order time-derivative, 
or to a "modified" form. For waiting time distributions which 
are not pure power laws one or the other form of the kinetic equation are shown to be 
advantageous, depending on whether the process slows down or accelerates in the course of time. 

\bigskip

\end{bf}

In his epochal 1905 work, see \cite{Einstein}, Einstein showed how the postulates of 
the kinetic theory of heat led to the conclusion that small but
macroscopic particles suspended in a fluid must perform an unceasing
motion. He obtained the laws governing this motion which he identified as Brownian motion. 
The motion in the absence of an external force was shown to be diffusive. 
The discussion of the stationary state in the
gravitational field led to quantitative relations from which, for example,
Avogadro's number could be obtained with a precision which at that time
couldn't be reached by other methods. The results of this and subsequent
works \cite{Einstein} made up Einstein's PhD thesis defended in Zurich in January 1906.
The approach proposed by Einstein included some of the modern approaches 
to stochastic processes, and stimulated and encouraged other
scientists such as Smoluchowski, Langevin and Planck who had their own interests 
in molecular motion. Thus, Smoluchowski in 1906
formulated a clearer probabilistic approach to the problem of diffusion based on what we
call now random walks, which, according to his own words, was more
straightforward and therefore simpler than Einstein's. 
Langevin in 1908 introduced an approach based on what nowadays is called
stochastic differential (or Langevin) equation, which was claimed to be
''infinitely simpler''. We should also mention that the widely used term ''random walk'' 
stemmed from a question put forward to readers of ''Nature'' by 
Carl Pearson in 1905 motivated by a biological problem.

Einstein's description of the motion of suspended particles was based on
three postulates:
\begin{itemize}

\item The motion of different particles is independent
(which is always true if their concentration is low enough), so that the
problem is essentially a one-particle one. 

\item There exists a time
interval $\tau $ such that the displacements of the same particle during
different $\tau $-intervals can be considered as independent (fast
decorrelation). 

\item There exists a mean squared displacement 
$\lambda ^{2}$ of the particle during such a $\tau$-interval . 

\end{itemize}

The second postulate guarantees the Markovian nature of displacement at times larger
than $\tau $, and the third one leads to the convergence of the
corresponding process to a normal diffusion (and, say, not to a L\'{e}vy
flight). 

The overall picture put forward by Einstein looks as follows: The state 
of the system (the coordinates of the Brownian particle) is sampled at 
time intervals of increment $\tau $ and the displacement during each
interval $\tau $ is chosen according to some probability density function (pdf) $p(s)$. 
This picture is essentially a random walk picture since considering the system 
only at times which are the multiples of $\tau $ leads to the notion of 
\textit{steps}: the pdf $p(s)$ plays the role of the distribution of step lengths,
and the waiting time until the next step is exactly $\tau $.

The formulation in terms of discrete jumps or steps is adequate in solids,
where one of the mechanisms of conduction is attributed to thermal
activation out of bound impurity states. However, here the idea of a fixed
waiting time $\tau $ between the jumps has to be abandoned, and the
waiting-time distribution $\psi (\tau )$ of the times between subsequent steps
has to be explicitly taken into account. This defines the continuous time random
walk process (CTRW) which was first introduced into mathematical physics
by Montroll and Weiss \cite{MontrollWeiss}, and applied to semiconductors in a
seminal work by Scher and Montroll \cite{ScherMontroll}. 
The \textit{decoupled} CTRW scheme assumes that the step lengths and the waiting times are independent 
random variables, chosen from corresponding distributions with probability densities  
$p(s)$ and $\psi(\tau)$. A simple argumentation based on an exponential 
form of the tail of density of impurity states and on the Arrhenius law leads to the
power-law form of the waiting-time distribution \cite{SSB}, 
\begin{equation}
\psi (\tau )\propto \tau ^{-1-\alpha }  \label{Eq1}
\end{equation}
where the exponent $\alpha $ is proportional to the temperature. 
Of particular interest is the case of $0<\alpha<1$ for which 
all moments of $\psi(\tau)$ diverge. This process has 
no characteristic time scale, a strong violation of the original 1905 picture,
and results in subdiffusion \cite{BouchaudGeorge,Blumofen,MetzlerKlafter,SKB}. The Markovian nature
of the process is violated as well, and replaced by a semi-Markovian character \cite{Feller}. In what
follows we use the scalar notation; however, no problems arise when taking
$x$ to denote the components of the particle's position in three-dimensional space. 

We note that the particle's displacement under a random walk process 
is given by the stochastic differential equation:
\begin{equation}
\frac{d}{dt}x=\sum a_{i}\delta (t-t_{i}),
\end{equation}
where $a_{i}$ gives the length and direction of the corresponding step
taking place at the time instant $t_{i}$. Coarse-graining this picture over
some typical time interval $\Delta t$ leads to a picture which corresponds to
the overdamped Langevin dynamics: if both the mean time between steps 
$\left\langle \tau \right\rangle =\left\langle t_{i+1}-t_{i}\right\rangle$
and the mean squared step length $\left\langle a_{i}^{2}\right\rangle$
exist, then averaging over the interval $\Delta t>>\left\langle \tau \right\rangle$ 
leads to appearance of a Gaussian noise $\xi (t)$ with intensity 
$\left\langle a_{i}^{2}\right\rangle/\tau $. In this coarse-grained sense 
one can also put down
\[
\dot{x}(t)=\xi (t).
\]
The noise is uncorrelated on intervals $\Delta t \gg \left\langle \tau \right\rangle$.
If the mean waiting time diverges, $0<\alpha<1$ in Eq.(\ref{Eq1}), no convergence to a 
Gaussian takes place for whatever long $t$ due to the existence of indefinitely long step-free 
intervals. The Langevin scheme for CTRWs with
power-law waiting time distribution assumes thus the introduction
of correlated non-Gaussian noise  \cite{Saichev} and loses its appealing properties as
intuitive description instrument. 

The original approach to CTRW based on the Laplace-Fourier-represen\-tation of the process can be
generalized to models showing correlations (like L\'{e}vy walks \cite{LevyWalks}) but, due
to the crucial assumption of spatial homogeneity, can hardly be used for the
description of systems in external fields unless these are homogeneous. On
the other hand, the uncorrelated CTRWs can be considered easily also for
spatially inhomogeneous situations, as long as this inhomogeneity concerns only step
lengths, and not waiting times. This is so, since this latter case corresponds 
exactly to the case of subordinated random processes: if almost all 
$\psi (\tau)$ (except at most the ones corresponding to boundary sites) are
the same, the dynamics of the system can be considered as a process
subordinated to simple random walks. To elucidate the situation let us start
from the system without physical boundaries.

Considering the situation not as a function of time, but rather as a function 
of the number of steps, we can easily convince ourselves that the particle's
displacement as a function of the number of steps is a discrete time random
walk, possibly with position-dependent step lengths. Let us consider the
number of steps as the internal, operational time governing the system's
evolution. In this operational time the evolution is given by the 
$x$-dependent probabilities for discrete jumps, or by probability densities 
for continuous step lengths. In what follows we discuss the continuous
situation, and describe the displacements by their pdf's; the discrete case
differs only terminologically. The role of the waiting times in this case
reduces to the fact that the actual number of steps made up to the time
instant $t$ fluctuates, so that the operational time is a random
function of the physical time $t$. The fact that this function is
monotonously nondecaying with $t$ and thus allows for causal ordering of the
events (i.e. is really a \emph{time}), is in general important,
but plays no role in the following considerations.
From the probabilistic point of view, the distributions for subordinated
processes at some physical time $t$ are mixtures of the corresponding
distributions of the underlying process at different operational times. For
example, for the pdfs of processes with continuous distributions of step
lengths one has
\begin{equation}
P(x,t)=\sum_{n}W(x,n)\chi_{n}(t)  \label{CTRW}
\end{equation}
where $W(x,n)$ is a probability distribution to find a random walker at
point $x$ after $n$ steps, and $\chi_{n}(t)$
is the probability to make exactly $n$ steps up to time $t$. In a classical
Scher-Montroll CTRW for which the waiting time distribution follows
Eq.(\ref{Eq1}) with $0<\alpha<1$, $\chi_{n}(t)$ corresponds to a random process in which $n$
typically grows sublinearly in $t$, $n(t) \sim t^\alpha$. One can say that the operational time 
is always in delay compared with the physical one, see Fig.1. Thus, the overall process is
subdiffusive. We note that some superdiffusive processes can also be considered as subordinated to random
walk; in this case the typical value of $n$ grows faster than linear in $t$ \cite{Sokolov1}.
In the present work we concentrate on the subdiffusive case.

\begin{figure}[tbh]
\begin{center}
\epsfxsize = 5in
\epsffile{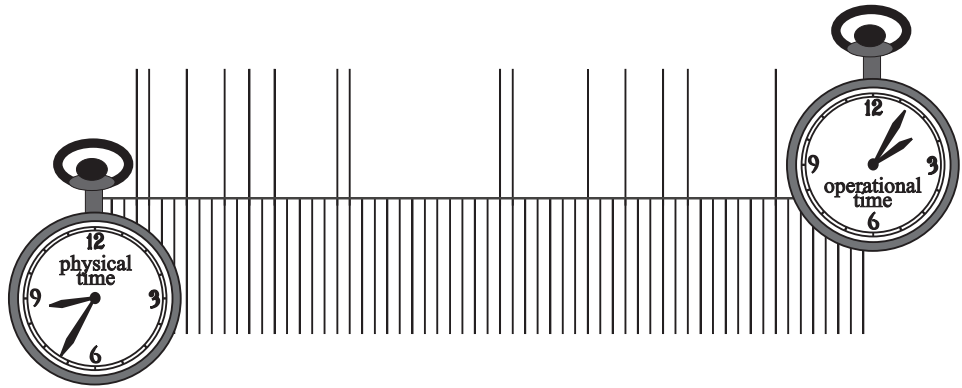}
\caption{The operational time of a CTRW process with a power-law
waiting time distribution between steps. The steps follow very irregularily
in physical time, and show step-free intervals on all time scales. On the
average, the operational time is in delay compared with the physical one.}
\end{center}
\end{figure}

In the continuous limit one can pass from a sum to an integral, changing the 
discrete variable $n$ to a continuous variable $\tau$, In this case 
\begin{equation}
P(x,t)=\int_{0}^{\infty }W(x,\tau )T(\tau ,t)d\tau .  \label{Sub1}
\end{equation}

The function $T(\tau ,t)$ for a decoupled subdiffusive CTRW can be easily
calculated, since the probabilities $\chi _{n}(t)$ are known 
\cite{MontrollWeiss,Blumofen}. Their Laplace transforms are given by 
\[
\tilde{\chi}_{0}(u) = \frac{1-\tilde{\psi}_{1}(u)}{u}
\]
and
\begin{equation}
\tilde{\chi}_{n}(u) = \tilde{\psi}_{1}(u)\tilde{\psi}^{n-1}(u)
\frac{1-\tilde{\psi}(u)}{u}\qquad (n \geq 1).
\label{chi}
\end{equation}
Here and in what follows the Laplace-transform of a function of time is
denoted by tilde and $\tilde{\psi}_1(u)$ is a Laplace transform of the forward waiting time
density of the first step $\psi_1(t)$, which might differ from $\psi(t)$ for all
subsequent steps. In what follows we take that the beginning of the time count
coincides with the preparation of the system ("zeroth" step), so that 
$\tilde{\psi}_{1}(u)=\tilde{\psi}(u)$. 
Moreover, this corresponds to the assumption that for $t=0$ one
also has $\tau=0$ so that the initial conditions for probability densities
$P(x,t)$ and $W(x,\tau)$ coincide.
For any $n$ we thus have \cite{MontrollWeiss,Blumofen} 
\begin{equation}
\tilde{\chi}_{n}(u)=\tilde{\psi}^{n}(u)\frac{1-\tilde{\psi}(u)}{u} =
\frac{1-\tilde{\psi}(u)}{u}\exp \left[ -n\ln \tilde{\psi}(u)\right] .
\end{equation}
Since $\tilde{\psi}(u)$ is a Laplace-transform of a pdf, 
it is continuous. In addition $\tilde{\psi}(0)=1$, which allows us to
introduce the function $\tilde{\phi}(u)=1-\tilde{\psi}(u)$ which is small for $u$ small
(corresponding to large $t$). 
In the time domain $\phi(t)$ is simply given by $\phi(t)=\delta(t)-\psi(t)$.
Expanding the logarithm we obtain 
\begin{equation}
\tilde{T}(n,u)\simeq \tilde{\chi}_{n}(u)\simeq \frac{\tilde{\phi}(u)}{u}\exp
\left[ -n\tilde{\phi}(u)\right] .
\label{Te}
\end{equation}
Applying Laplace transform to both sides of Eq.(\ref{Sub1}), 
\begin{eqnarray}
\tilde{P}(x,u) &=&\int_{0}^{\infty }W(x,\tau )\tilde{T}(\tau ,u)d\tau 
\nonumber \\
&=&\int_{0}^{\infty }W(x,\tau )\frac{\tilde{\phi}(u)}{u}\exp \left[ -\tau 
\tilde{\phi}(u)\right] d\tau  \nonumber \\
&=&\frac{\phi (u)}{u}\int_{0}^{\infty }W(x,\tau )\exp \left[ -\tau 
\tilde{\phi}(u)\right] d\tau  \nonumber \\
&\equiv &\frac{\tilde{\phi}(u)}{u}\tilde{W}(x,\tilde{\phi}(u)),  \label{Sula}
\end{eqnarray}
i.e. $\tilde{P}(x,u)$ can be obtained from $\tilde{W}(x,u)$ through a change 
of variable. The particular case of $\tilde{\phi}(u) \sim u^\alpha$ is considered in
\cite{Metzleretal}. One can make a step
further and obtain the equation governing the behavior of $P(x,t)$, if the
equation giving the ''time''-evolution of $W(x,\tau )$ is known. Let us
assume that $W(x,\tau )$ satisfies the following equation 
\begin{equation}
\frac{d}{d\tau }W(x,\tau )=\mathcal{L}_{x}W(x,\tau),  \label{Linea}
\end{equation}
where the linear operator $\mathcal{L}_{x}$ acting on the $x$-variable of $W$
might be discrete as in the master equation, or continuous as in the
diffusion equation with 
\[
\mathcal{L}_{x}=K \frac{\partial^2}{\partial x^2}
\]
or in a Fokker-Planck equation
\[ 
\mathcal{L}_{x}=\mathcal{L}_{\mathrm{FPE}}= K \frac{\partial^2}{\partial x^2}-\mu 
f(x)\frac{\partial}{\partial x},
\]
where $K$ is the diffusion coefficient, $\mu$ is the mobility, and $f(x)$ the external force.
It is only important that it does not depend on any of
the time variables of the problem, $t$ or $\tau $ (the last assumption can
be relaxed, see \cite{Stanislavsky}). The Laplace-transform of Eq.(\ref{Linea}) reads 
\begin{equation}
s\tilde{W}(x,s)-W(x,0)=\mathcal{L}_{x}\tilde{W}(x,s)
\label{LineaLap}
\end{equation}
where $s$ is the Laplace variable. Taking $s=\tilde{\phi}(u)$ we get 
\begin{equation}
\tilde{\phi}(u)\tilde{W}(x,\tilde{\phi}(u))-W(x,0)=\mathcal{L}_{x}
\tilde{W}(x,\tilde{\phi}(u)),
\end{equation}
and multiplying both sides by $\tilde{\phi}(u)/u$ 
\footnote{The function $\tilde{\psi}(u)$ is a Laplace transform of a pdf and therefore
is completely monotonous. This means that it is positive and monotonously
non-growing. Moreover, $\tilde{\psi}(u)=1$ so that 
$\tilde{\phi}(u)$ is monotonously nondecaying and $\tilde{\phi}(0)=0$. For all
distributions except for $\delta (t)$ it is monotonously growing and thus
possesses no zeros except for one for $u=0$ (which corresponds to 
$t=\infty $). Therefore all divisions or multiplications discussed in the text are
essentially harmless operations.} 
we obtain the equation for $\tilde{P}(x,u)$
\begin{equation}
\tilde{\phi}(u)\tilde{P}(x,u)-\frac{\tilde{\phi}(u)}{u}W(x,0)=\mathcal{L}_{x}\tilde{P}(x,u).  
\label{KinEq}
\end{equation}
From this equation different forms of the corresponding kinetic equation in the
time domain can be obtained. These forms are essentially equivalent,
however, depending on the analytic properties of the function $\phi (u)$,
one form or another can be preferable for practical applications.

Returning to the time-domain we get from Eq.( \ref{KinEq})
the following integro-differential equation:
\begin{equation}
\int_{0}^{t}\phi (t-t^{\prime })P(x,t^{\prime })dt^{\prime
}-W(x,0)\int_{0}^{t}\phi (t^{\prime })dt^{\prime }=\mathcal{L}_{x}P(x,t),
\label{KinEq1}
\end{equation}
where the function $\phi (t)$ is the inverse Laplace transform of the
function $\tilde{\phi}(u)$.  Inserting $\phi (t)=\delta (t)-\psi (t)$,
performing partial integration in the convolution term, and using the fact 
that $P(x,0)=W(x,0)$, one arrives at
\begin{equation}
\int_{0}^{t}C(t-t^{\prime })\frac{\partial}{\partial t^{\prime}} 
P(x,t^{\prime })dt^{\prime}=\mathcal{L}_{x}P(x,t).
\label{Cap1}
\end{equation}
The integral kernel $C(t)$ here is connected to the cumulative
distribution function of waiting times, 
$C(t)=1-F(t)=\int_t^\infty \psi(t^{\prime}) d t^{\prime}$, i.e. it is
the probability to make no step up to time $t$. This form of the kinetic equation
will be termed here as a generalized Caputo form, for the reasons which will be clear
below. 

The kinetic equation, Eq.(\ref{Cap1}), has a clear mathematical meaning, and nonnegative
solutions (subordinated to those of the Markovian evolution equation, Eq.(\ref{LineaLap}))
for all nonnegative, monotonously decaying kernels $C(t)$.  
Essentially one can even say that all equations of the form of Eq.(\ref{Cap1}) 
whose kernels are monotonously non-growing, nonnegative functions
describe continuous limits of decoupled CTRW. 

A simple example of the exponential waiting time pdf
\begin{equation}
\psi(\tau)=\lambda \exp(-\lambda \tau)
\label{MarkProc}
\end{equation}
leads to $C(t)=\exp(-\lambda t)$. For $t \gg \lambda^{-1}$ the kernel $C(t)$ may be 
approximated by a $\delta$-function,  $C(t)\simeq \lambda^{-1} \delta(t)$, 
and the integration over $t'$ leads to
\begin{equation}
\frac{\partial}{\partial t} P(x,t)=\lambda \mathcal{L}_{x}P(x,t),
\end{equation}
corresponding to a simple time-\textit{scale} change by a factor of $\lambda$. 
The process is back to Markovian.

In case of power-law distributions $\psi(t)$, Eq.(\ref{Eq1}),
the left hand side (lhs) of Eq.(\ref{Cap1}) defines \cite{Caputo,Chech1} an interesting mathematical 
object, namely a fractional Caputo derivative. The Caputo derivative is an operator of the form 
\begin{equation}
~_{0}D_{*t}^{\alpha } f(t)=~_{0}I_{t}^{1-\alpha }\frac{d}{dt}f(t)=\frac{1}
{\Gamma (1-\alpha )}\int_{0}^{t}d\tau (t-\tau )^{-\alpha }\frac{d}{d\tau }
f(\tau ).
\end{equation}
Here $~_{0}I_{t}^\beta$ is a fractional integral operator defined through
\begin{equation}
~_{0}I_{t}^{\beta} g(t) = \frac{1}{\Gamma(\beta )} \int_0^t (t-\tau)^{\beta-1} g(\tau) d\tau,
\label{FracInt}
\end{equation}
$0<\beta<1$. The Laplace representation of this operator corresponds to multiplication by $u^{-\beta}$. 
Thus, the Laplace representation of $~_{0}D_{*t}^{\alpha }$ is
$~_{0}D_{*t}^{\alpha } f(t) \risingdotseq u^{-1+\alpha }(uf(u)-f(0))$, exactly the form in the
lhs of Eq.(\ref{KinEq}) for $\phi (u)=u^{\alpha }$. This equation
corresponds to a ''normal'' form of the kinetic equation \cite{ActaPhysica} with a fractional
operator instead of first-order time-derivative on the lhs: 
\begin{equation}
~_{0}D_{*t}^{\alpha }P(x,t)=\mathcal{L}_{x}P(x,t).
\label{Caputo}
\end{equation}

Integro-differential equations of the type of Eq.(\ref{Cap1}) do not have the 
form one typically uses. We now obtain a conjugated form, resembling the 
generalized Fokker-Planck equation \cite{Zwanzig}. Multiplying both sides of Eq.(\ref{KinEq}) by 
$u/\tilde{\phi}(u)$ we get 
\begin{equation}
u\tilde{P}(x,u)-W(x,0)=u\frac{1}{\tilde{\phi}(u)}\mathcal{L}_{x}\tilde{P}(x,u),
\end{equation}
which is the Laplace-transform of a generalized kinetic equation in the form 
\begin{equation}
\frac{\partial }{\partial t}P(x,t)=\frac{d}{dt}\int_{0}^{t}\Phi (t-t^{\prime
})\mathcal{L}_{x}\tilde{P}(x,t^{\prime })dt^{\prime }.
\label{modified}
\end{equation}
The lower integration limit follows from the fact that the Laplace representation of 
the corresponding integro-differential operator coincides with  $u/\tilde{\phi} (u)$ only if 
the integral term vanishes at the initial condition. The main difference when compared 
to the typically used forms \cite{Kenkre} is the additional time derivative in front of the integral.
Note that the memory kernel $\Phi (t)$ of the integro-differential operator 
has a clear physical meaning.
Let us expand
\begin{equation}
\frac{1}{\tilde{\phi}(u)}=\frac{1}{1-\tilde{\psi}(u)}=1+\tilde{\psi}(u)+
\tilde{\psi}^{2}(u)+\tilde{\psi}^{3}(u)+...
\end{equation}
so that
\begin{equation}
\Phi (t)=\delta (t)+\psi +\psi *\psi (t)+\psi *\psi *\psi (t)+...
\end{equation}
where $\psi *...*\psi (t)=\int dt_{1}...\int dt_{n}\psi (t_{1})\psi
(t_{2})...\psi (t-t_{n})$ is the $n$-th convolution of the pdf $\psi (t)$
with itself, which gives us the pdf that the $n$-th step of the CTRW takes
place at time $t$. The sums of these probabilities gives us the 
\textit{density of steps}: $\Phi (t)dt$ is the mean number of steps taken in CTRW
in the time interval between $t$ and $t+dt$, which makes the overall picture
given by the equation rather transparent. Thus, the integral kernels for the 
equations describing CTRWs have to be nonnegative. 

Contrary to Eq.(\ref{Cap1}), which is a valid
kinetic equation for any nonnegative decaying kernel, the analytical
properties of Eq.(\ref{modified}) are more complex: some kernels do not 
correspond to any decoupled CTRW, others even do not warrant
the non-negativity of solutions (i.e. are "dangerous", see Ref.\cite{Sokolov2}).
Let us show that a necessary condition for Eq.(\ref{modified}) to describe the decoupled CTRW
is the divergence of the integral $\int_0^\infty \Phi(t)dt$.
To see this it is enough to note that for small $u$ one has $\phi(u) \rightarrow 0$, 
so that the Laplace transform  $\Phi(u)$ of the kernel $\Phi(t)$ has to diverge for small $u$;
otherwise the normalization of the waiting-time density is violated.
This divergence means that the integral of the kernel over the time must also diverge.
Equations with integrable kernels do not describe any decoupled CTRW-like process, although
they might appear in a context of coupled motion \cite{Yossi}. 
As an example we see that the exponential kernel $\Phi(t)=\exp(-t/t_0)$ 
(for which Eq.(\ref{modified}) with a Fokker-Planck operator 
$\mathcal{L}_{x}$ reduces to a kind of a telegrapher's equation) can never 
appear as a true kernel of the decoupled problem \cite{Sokolov2}. 

For a Poisson process characterized 
by the waiting-time pdf, Eq.(\ref{MarkProc}) 
the density of steps tends to a constant $\Phi(t) \rightarrow \lambda$, 
so that applying the integro-differential operator is equivalent to a 
multiplication by a constant, which corresponds only to the change of the time-scale.

For the power-law waiting-time density, Eq.(\ref{Eq1}), we have 
$\tilde{\phi}(u)\propto u^{\alpha }$, so that $\tilde{\Phi}(u)\propto u^{-\alpha }\,$ so that
$\Phi (t)\propto t^{-1+\alpha }$ is a kernel of a fractional integral
operator $_{0}I_{t}^{\alpha }$ defined by Eq.(\ref{FracInt}).
The integro-differential operator in Eq.(\ref{modified}) corresponds to the operator of the  
Riemann-Liouville fractional derivative of the order $1-\alpha $, see e.g. \cite{MetzlerKlafter,SKB}. 
This is what we call the ''modified'' 
form of the fractional Fokker-Planck equation introducing an additional Riemann-Liouville
fractional derivative on the right hand side (rhs) of the equation
which now reads 
\begin{equation}
\frac{\partial }{\partial t}P(x,t)=~_{0}D_{t}^{1-\alpha }\mathcal{L}_{x}
\tilde{P}(x,t),  \label{DEq1}
\end{equation}
where the operator $_{0}D_{t}^{\beta }$ is defined through: 
\begin{equation}
_{0}D_{t}^{\beta }f=\frac{d}{dt}~_{0}I_{t}^{1-\beta }f=\frac{1}{\Gamma
(1-\beta )}\frac{d}{dt}\int_{0}^{t}d\tau (t-\tau )^{-1+\beta }f(\tau ).
\end{equation}

Pure fractional kinetic equations such as Eq.(\ref{Caputo}) and Eq.(\ref{DEq1}) 
are not the only ''reasonable'' forms. 
In general both forms, the ''normal'' and the ''modified'' one, are
equivalent. However, for waiting-time distributions which asymptotically
are not pure power-laws one or the other might be preferable for practical
applications, depending on the properties of the waiting-time distribution.
Special cases are kinetic equations with a distributed-order 
Caputo derivative \cite{Chech1,Chech2} 
\begin{equation}
\int_{0}^{1}d\alpha w_1(\alpha)~_{0}D_{*t}^{1-\alpha }P(x,t) = \mathcal{L}_{x}P(x,t),
\end{equation}
and with a distributed-order Riemann-Liouville derivative  \cite{ActaPhysica}
\begin{equation}
\frac{\partial }{\partial t}P(x,t)=\int_{0}^{1}d\alpha w_2(\alpha
)~_{0}D_{t}^{1-\alpha }\mathcal{L}_{x}\tilde{P}(x,t),
\end{equation}
where $w_i(\alpha )$ are some weight function. 

The "normal" form with distributed-order Caputo derivative is well fitted for
describing processes getting more anomalous in the course of the time (retarding
subdiffuion, as exemplified by crossover models of Refs.\cite{Chech1,Chech2} and by 
Sinai-like diffusion processes \cite{ChechkinKS}), while the
"modified" form with a distributed-order Riemann-Liouville derivative describes
the processes which in the course of the time get less anomalous (accelerating
subdiffusion). Let us demonstrate these statements with two examples of different 
waiting-time distributions corresponding to crossover behaviors. 
Here it is easier to start from the asymptotic forms
of cumulative distribution functions $\Psi(t)=1-C(t)=\int_0^t \psi(t')dt'$,
for $t \gg 1$: 
\begin{equation}
\Psi _{1}(t)\simeq 1-\frac{a}{t^{\alpha }}-\frac{b}{t^{\beta }}
\end{equation}
and 
\begin{equation}
\Psi _{2}(t)\simeq 1-\frac{1}{ct^{\alpha }+dt^{\beta }}
\end{equation}
corresponding to an arithmetic and a harmonic means of
the two power-laws with powers $\alpha $ and $\beta $, $\alpha <\beta $.
Both waiting-time distributions correspond to a crossover between two
regimes each governed by a different power. In the case of $\Psi
_{1}$ short times are dominated by the larger power $\beta $ and
longer times are dominated by smaller power $\alpha $. The crossover between
the two regimes takes place at $t\simeq t_{c}=\left( b/a\right) ^{1/(\beta
-\alpha )}$. In the case of $\Psi _{2}$ the behavior at short times is
dominated by the smaller power $\alpha $ while the behavior at long times is
dominated by the larger power $\beta $. The crossover takes place at 
$t_{c}\simeq (c/d)^{1/(\beta -\alpha )}$. The long-time asymptotics of the
corresponding waiting-time pdf's are 
\begin{equation}
\psi _{1}(t)\simeq \frac{a\alpha }{t^{\alpha +1}}+\frac{b\beta }{t^{\beta +1}}
\label{psi1}
\end{equation}
which behaves as $\psi _{2}(t)\simeq \frac{a\alpha }{t^{\alpha +1}}$ for 
$t\lesssim t_{c}$ and as $\psi _{2}(t)\simeq \frac{b\beta }{t^{\beta +1}}$
for $t\gtrsim t_{c}$, and 
\begin{equation}
\psi _{2}(t)=\frac{c\alpha t^{\alpha -1}+d\beta t^{\beta -1}}{\left(
ct^{\alpha }+dt^{\beta }\right) ^{2}}
\label{psi2}
\end{equation}
which behaves as $\psi _{2}(t)\simeq \frac{\beta }{dt^{\beta +1}}$ for 
$t\lesssim t_{c}$ and as $\psi _{2}(t)\simeq \frac{\alpha }{ct^{\alpha +1}}$
for $t\gtrsim t_{c}$. A particular example of such pdf's is given in Fig.2.

\begin{figure}[tbh]
\begin{center}
\epsfxsize = 3.5in
\epsffile{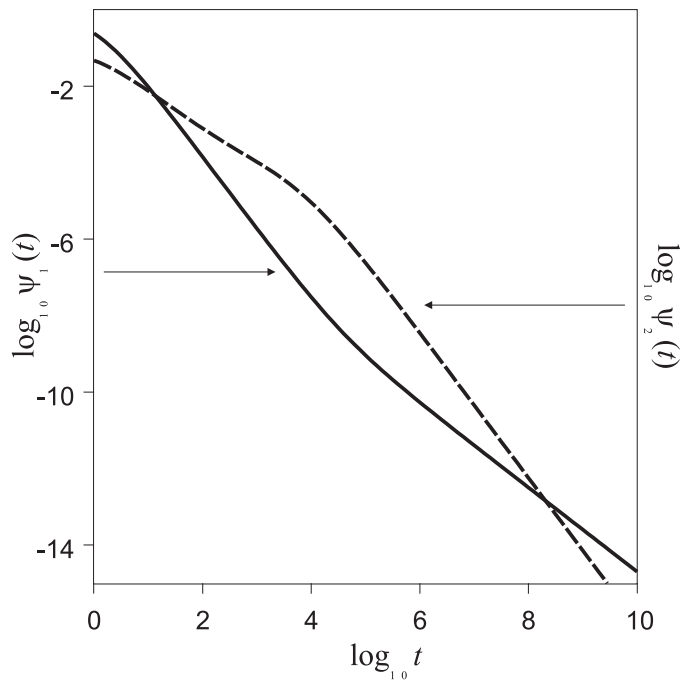}
\caption{The probability density functions
$\psi_{1}(t)$, Eq.(\ref{psi1}) (full line), and $\psi _{2}(t)$, Eq.(\ref{psi2}) (dashed line).
The parameters are $\alpha=0.1$, $\beta=0.9$, $a=d=10^{-3}$, $b=c=1-10^{-3}$.}
\end{center}
\end{figure}

Let us discuss first the types of the CTRW processes described by such waiting 
time distributions. To do this we discuss the mean number of steps (mean 
operational time) as a function of $t$. This quantity is of immediate
importance since it is proportional to the mean squared displacement in a
free diffusion or in a homogeneous external field. 
To obtain this value we note that from
Eq.(\ref{chi}) it follows that  
\begin{equation}
\tilde{n}(u)=\sum_{n=0}^\infty n \tilde{\chi}_n(u) \simeq \frac{\tilde{\psi}(u)}
{u\left[ 1-\tilde{\psi}(u)\right] }\simeq \frac{1}{u}\frac{1}{\tilde{\phi}(u)},
\end{equation}
which corresponds to the time-integral of the function $\Phi(u)$. 
Let us calculate the functions $\Phi(u)$ for both our distributions. 
For the case of $\Psi_1$ we get: 
\begin{equation}
\tilde{\psi}_{1}(u)=1-\left( Au^{\alpha }+Bu^{\beta }\right) 
\end{equation}
so that the function $\Phi (u)$ is
\begin{equation}
\tilde{\Phi}_{1}(u)=\frac{1}{1-\tilde{\psi}_{1}(u)}=\frac{1}{Au^{\alpha
}+Bu^{\beta }}
\label{Phii1}
\end{equation}
and 
\begin{equation}
\tilde{n}_{1}(u)=\frac{1}{u}\frac{1}{Au^{\alpha }+Bu^{\beta }}
\end{equation}
and use the Tauberian theorem, stating that the Laplace-transform of a function 
$f(t)\simeq t^{-\beta }L(t)$, where $L(t)$ is a slowly changing function of
the time (for which $\lim_{t\rightarrow \infty }\left( L(kt)/L(t)\right) =1$
for any positive constant $k$) corresponds in the Laplace representation
for $u\rightarrow 0$ to a function  
$\tilde{f}(u)\simeq u^{\beta -1}\Gamma (1-\beta )L(1/u)$. 
In our case
\begin{equation}
\tilde{n}_{1}(u)=\frac{1}{Au^{\alpha +1}}\frac{1}{1+(B/A)u^{\beta -\alpha }}=
\frac{1}{Au^{\alpha +1}}\frac{1}{1+(B/A)\left( 1/u\right) ^{\alpha -\beta }}
\end{equation}
is a product of a power-law and of a slowly changing function of the
variable $1/u$ and thus is a Laplace transform of a function which asymptotically
behaves as 
\begin{eqnarray}
n_{1}(t) &\simeq &\frac{1}{A\Gamma (\alpha +1)}t^{\alpha }\frac{1}
{1+(B/A)t^{\alpha -\beta }}  \nonumber \\
&=&\frac{1}{\Gamma (\alpha +1)}\frac{1}{At^{-\alpha }+Bt^{-\beta }}.
\label{N1}
\end{eqnarray}

In the case of $\Psi_2$ the function $\tilde{\Phi}_2(u)=1/\tilde{\phi}_2(u)$ 
can be also obtained using the Tauberian Theorem. To do this we note that
the function $\phi_{2}(t)$ is a derivative of $C_{2}(t)$, and therefore
its Laplace-transform is 
\begin{equation}
\tilde{\phi}_{2}(u)=u\tilde{C}_{2}(u).
\end{equation}
On the other hand, $\tilde{C}_{2}(u)$ is the Laplace transform of the function
\begin{equation}
C_{2}(t) = \frac{1}{ct^{\alpha }+dt^{\beta }}=\frac{d^{-1}}{t^{\beta }}
\left( \frac{1}{1+\left( c/d\right) t^{\alpha -\beta }}\right)
\end{equation}
being a product of a power-law and a slowly changing function. 
Therefore
\begin{equation}
\tilde{C}_{2}(u)\simeq \Gamma (1-\beta )d^{-1}u^{\beta -1}L(1/u)=\Gamma
(1-\beta )d^{-1}u^{\beta -1}\frac{1}{1+(c/d)u^{\beta -\alpha }}.
\label{CTaub}
\end{equation}
The Laplace transforms, $\tilde{\Phi}_2(u)$ and  $\tilde{n}_{2}(u)$ for this 
case read
\begin{equation}
\tilde{\Phi}_{2}(u)\simeq \frac{1}{u}\Phi _{2}(u)=\frac{1}{\Gamma (1-\beta )}%
\left[ du^{-1-\beta }+cu^{-1-\alpha }\right] 
\label{Phii2}
\end{equation}
and
\begin{equation}
\tilde{n}_{2}(u)\simeq \frac{1}{u}\Phi _{2}(u)=\frac{1}{\Gamma (1-\beta )}
\left[ du^{-1-\beta }+cu^{-1-\alpha }\right],
\end{equation}
so that
\begin{equation}
n_{2}(t)=\frac{1}{\Gamma (1-\beta )}\left[ \frac{d}{\Gamma (1+\beta )}
t^{\beta }+\frac{c}{\Gamma (1+\alpha )}t^{\alpha }\right]. 
\label{N2}
\end{equation}
Since all Gamma-functions in Eqs.(\ref{N1}) and (\ref{N2}) are of the order of unity, the
typical crossover times between the short- and long-time asymptotic behavior for $n(t)$ 
are the same as in the corresponding waiting times. 

Let us now compare the forms of the corresponding generalized diffusion and
(Fokker-Planck) equations. To obtain these generalized equations we have to calculate 
the kernels in Eqs.(\ref{Cap1}) and (\ref{modified}). For the first function, the behavior of $C$, follows
trivially: 
\begin{equation}
C_{1}(t)=\frac{a}{t^{\alpha }}+\frac{b}{t^{\beta }}.
\end{equation}
Taking into account the definitions of the corresponding fractional
operators, we see that the generalized diffusion equation has a simple form
with two fractional Caputo derivatives on the lhs: 
\begin{equation}
\left( A_1~_{0}D_{*t}^{\alpha }+B_1~_{0}D_{*t}^{\beta }\right) P(x,t)=\mathcal{L}_x P(x,t)
\end{equation}
with $A_1=\Gamma (1-\alpha )a$ and $B_1=\Gamma (1-\beta )b$. Quite opposite, the
function $\Psi _{2}(t)$ does not correspond to any simple form of the
equation. The kernel $C_2(t)$ does not possess a form resembling
any fractional operator, except in its far asymptotics $C(t)\propto
t^{-\beta }$. This confirms our statement that the ''normal'' form, 
reducing in our case to the form with two Caputo derivatives of different orders, 
is advantageous when describing the processes which get more and
more anomalous (in our case: slower and slower) in the course of time.

Let us now turn to the ''modified'' form. Using the 
Laplace-transforms of the corresponding $\Phi$-functions, Eqs.(\ref{Phii1}) and (\ref{Phii2}) 
we see that $\tilde{\Phi}_1(u)$ does not correspond 
to a Laplace transform of any simple fractional operator. On the other hand, the
Laplace transform of the function $\Phi _{2}$ corresponds to the Laplace-transform 
of the sum of two kernels of Riemann-Liouville fractional integrals of the orders of 
$\alpha $ and $\beta $. Therefore the corresponding "modified" equation 
has two Riemann-Liouville fractional derivatives on the rhs 
\begin{equation}
\frac{\partial }{\partial t}P(x,t)=\left( A_2~_{0}D_{t}^{1-\alpha
}+B_2~_{0}D_{t}^{1-\beta }\right) \mathcal{L}_x P(x,t)
\end{equation}
with $A_2=c/\Gamma (1-\beta )$ and $B_2=d/\Gamma (1-\beta )$. In general, the
''modified'' form is advantageous when describing processes which get less anomalous 
(here less subdiffusive) in the course of time.

In summary, we have considered a decoupled CTRW model, in which
the directions and the lenghts of steps, and the waiting times at sites are mutually
independent random variables. Such a CTRW can be considered as a random process 
subordinated to simple random walks. Using the idea of temporal subordination, 
we derive two different, but equivalent, forms of kinetic equations describing 
their long-time behavior. The corresponding equations the integro-differential ones,
and they reduce to the fractional equations with a Caputo- and a Riemenn-Liouville 
derivatives for waiting-time distributions characterized by the asymptotic power-law
pdfs. We discuss the advantages of each of the representation by considering two different
pdf's showing a crossover from one power-law behavior to the other. 
The corresponding kinetic equations might then take a form of
fractional equations with Caputo or with Riemann-Liouville derivatives of
distributed order. 

The authors are thankful to A.V. Chechkin, R. Metzler and R. Gorenflo for valuable discussions.
IMS gratefully acknowledges the support by the Fonds der Chemischen Industrie.

\end{document}